\documentclass[10pt]{article} 
\usepackage{opex3}
\usepackage{amsmath}
\usepackage[usenames,dvipsnames]{xcolor}
\usepackage{float}

\begin{document}
\title{Shaping the evanescent field of optical nanofibers for cold atom trapping}
\author{C.F.~Phelan, T.~Hennessy and Th.~Busch}

\address{University College Cork, Cork, Ireland \\ OIST (Okinawa Institute of
  Science and Technology), 1919-1 Tancha, Onna-son,
  Okinawa, Japan 904-0495}

\email{ciaran.phelan@oist.jp} 

\begin{abstract}
We investigate trapping geometries for cold, neutral atoms that can be created in the evanescent field of a tapered optical fibre by combining the fundamental mode with one of the next lowest possible modes, namely the ${HE_{21}}$ mode. Counter propagating red-detuned ${HE_{21}}$ modes are combined with a blue-detuned ${HE_{11}}$ fundamental mode to form a potential in the shape of four intertwined spirals. By changing the polarization from circular to linear in each of the two counter-propagating {${HE_{21}}$} modes simultaneously the 4-helix configuration can be transformed into a lattice configuration. The modification to the 4-helix configuration due to unwanted excitation of the the $TE_{01}$ and $TM_{01}$ modes is also discussed.
\end{abstract}


%%%%%%%%%%%%%%%%%%%%%%% References %%%%%%%%%%%%%%%%%%%%%%%%%

%%%%%%%%%%%%%%%%%%%%%%%%%%  body  %%%%%%%%%%%%%%%%%%%%%%%%%%

\section{Introduction}
\label{sec:intro}

A sub-wavelength diameter optical fibre can be produced by heating and pulling a standard telecommunications fibre so that its waist diameter reduces from about a hundred micrometers to a few hundred nanometers \cite{Tong:03,Ward:06,Yariv:85}. These tapered nanofibres have many uses and one of the most prominent is in the study of atomic samples and their optical properties at ultracold temperatures. By locally probing atomic fluorescence emitted from atoms trapped in a magneto-optical trap with high efficiency, they have been used to estimate the size and profile of the atomic cloud and other trap parameters \cite{Morrissey:09} and by recording the fluorescence spectrum of a small number of atoms close to the fibre surface, the effects of the short range van der Waals interaction has been investigated. \cite{Nayak:07}. 

When light travelling in such a fiber arrives at the tapered waist, the fiber diameter is smaller than the wavelength of the propagating light and  it can no longer be confined in the fiber. A considerable fraction of the power propagates outside the surface boundary in the form of an evanescent field. This presents a novel strategy for trapping and  guiding atoms near the fiber surface. The evanescent field represents an intensity gradient to a nearby atom, which leads to a dipole force that is either attractive or repulsive depending on whether the guided mode is red- or blue-detuned relative to the atom's dominant transition frequency. 

If both a red- and blue-detuned fundamental mode are present in the fiber at the same time, the differing decay lengths of their respective evanescent fields result in a combined optical potential with a minimum at some distance from the fiber surface \cite{Ovichnikov:91,LeKien:04}. This technique has been experimentally demonstrated for cold caesium atoms \cite{Vetsch:10}. The two-color trap can be made state insensitive by using red- and blue-detuned magic wavelengths \cite{LeKien:05,Goban:12} and vector light shifts due to the elliptic polarization of the nanofiber modes can be removed by introducing counter propagating beams\cite{Lacroute:12,Goban:12}. The evanescent field around a tapered optical fiber offers a strategy for gaining near-field access to atoms and characterising potential trapping geometries\cite{Sague:08}, and exploring possibilities for their use in more involved settings \cite{Hennessy:12} is becoming a very active research area. The advantages of using fibers that are slightly bigger than single-mode fibers has also become a subject of great interest \cite{Sague:08, Masalov:13}.

Here we present a scheme for creating helicoidal potentials by combining a counter-propagating higher mode with a blue-detuned fundamental mode. These spiralling geometries can be produced by fixing the blue-detuned portion of the evanescent field in a cylindrically symmetric configuration while modifying the red-detuned portion of the evanescent field. The creation of helical potentials in free space has previously been considered \cite{Okulov:12}, where they were generated by counter-propagating Laguerre Gaussian beams with counter-directed orbital angular momenta. One application of these spiralling potentials is, for example, the measurement of quantized rotations in atomic clouds \cite{Okulov:12_2}.

In the next two sections we will first briefly introduce and review the modes that can propagate in an optical nanofiber and discuss the trapping geometries that can be constructed using the fundamental modes. In Section \ref{sec:Traps} we present different trapping geometries which can be created by exciting a specific higher order mode in the fiber and in Section \ref{sec:powertransfer} we consider modification of the trapping geometry due to unintended excitation of nearby modes. Finally we conclude in Section \ref{sec:Conclusions}.

\section{Trapping with the first fundamental mode}
\label{sec:FiberModes}
An optical nanofiber can be thought of as a very thin optical fiber with a cylindrical silica core of radius $a$ and refractive index $n_1=1.452$ and an infinite vacuum clad of refractive index $n_2=1$.
It can be created by heating and pulling commercial grade optical fiber and tapering it down such that the refractive indices that determine the guiding properties of the fiber are that of the original silica cladding and that of the surrounding vacuum. Such a thin fiber can only support a finite number of modes and the permitted propagation constants, $\beta$, can be determined numerically \cite{Yariv:85}.
The field distributions associated with these modes can be found by solving Maxwell's equations and here we will just give the expressions for the evanescent part of the electric field as they will be used throughout this paper. In cylindrical polar co-ordinates $\{r,\phi,z\}$, these are given by \cite{Yariv:85}
\begin{subequations}
\begin{align}
  E_r(r,\phi,z)=&\;A\;\frac{J_l(ha)}{K_l(qa)}\frac{i\beta}{q^2}
                 \left[K_l^{'}(qr)+B\frac{i\omega\mu l}{\beta r}K_l(qr)\right] 
                 e^{i(\omega t+l\phi -\beta z)},\\
  E_\phi(r,\phi,z)=&\;A\;\frac{J_l(ha)}{K_l(qa)}\frac{i\beta}{q^2} 
           \left[\frac{il}{r} K_l(qr) -B\frac{\omega\mu}{\beta}K_l^{'}(qr)\right]
           e^{i(\omega t+l\phi -\beta z)},\\
  E_z(r,\phi,z)=&\;A\;\frac{J_l(ha)}{K_l(qa)}\ K_l(qr)e^{i(\omega t+l\phi -\beta z)},
\end{align}
  \label{eq:EvFields}
\end{subequations}
where the $J_l$ are Bessel functions of the first kind and the  $K_l$ are modified Bessel functions of the second kind. The dash denotes the derivative with respect to the argument of the Bessel function. The constant $B$ is given by $B=\frac{i\beta l}{\omega\mu}\left(\frac{1}{q^2 a^2}+\frac{1}{h^2a^2}\right)\left(\frac{J_l^{'}(ha)}{ha J_l(ha)}+\frac{K_l^{'}(qa)}{qa k_l(qa)}\right)$ and $A$ determines the power in a given mode. The parameter $q$, which is the reciprocal of the decay length of the evanescent field, is defined by $q=\sqrt{\beta^2-k_0^2 n_{2}^2}$ and $h=\sqrt{k_0^2 n_1^2-\beta^2}$.  The azimuthal index $l$ counts the number of $2\pi$ phase changes of each component in a circle around the fiber axis and its sign determines the polarization of the mode.  

The fields described by Eqns.~\eqref{eq:EvFields}  are not in general transversely polarized due to the $z-$component which, for a set frequency and decreasing fiber radius, increases in magnitude (relative to the transverse components). The transverse parts of the light beams described by these formulae are circularly polarized where the sign of the index $l$ gives the handedness of the polarization. Linearly polarized solutions can be obtained by taking superpositions of the circularly polarized beams with equal amplitude and opposite handedness.

The finite number of modes that can be supported by the fiber is determined by the ratio of the fiber radius to the wavelength of the propagating light. 
Under the constraint $V\equiv k_0 a\sqrt{n_1^2-n_2^2}<2.405$, the fiber can support only one mode, namely the fundamental mode HE$_{11}$.  Here $k_0$ is the free space wave-number. The $HE_{11}$ mode has a Gaussian intensity profile in the fiber and the shape of its evanescent field depends on three fundamental parameters of the system, the wavelength of the light, the fiber radius and the refractive index of the medium. 
If the light in the $HE_{11}$ mode is circularly polarized the intensity is azimuthally uniform. If the light is quasi-linearly polarised this uniformity is broken and two intensity maxima appear at opposite sides of the fiber \cite{LeKien:04}. The intermediate cases of elliptical polarization interpolate between these two scenarios. The evanescent field decays faster for shorter wavelengths.

An atom interacting with the evanescent field sees a dipole potential of the form $U=-\frac{1}{4}\alpha (\textbf{E}^{*}\textbf{E})$, where $\alpha$ is the atomic polarizability. It can be calculated from Eqs.~\ref{eq:EvFields}, and any intensity gradient results in a force whose sign is given by the detuning of the light field with respect to the atoms's dominant frequency. 

\begin{figure}[h]
   \centering
   \includegraphics[width=5in]{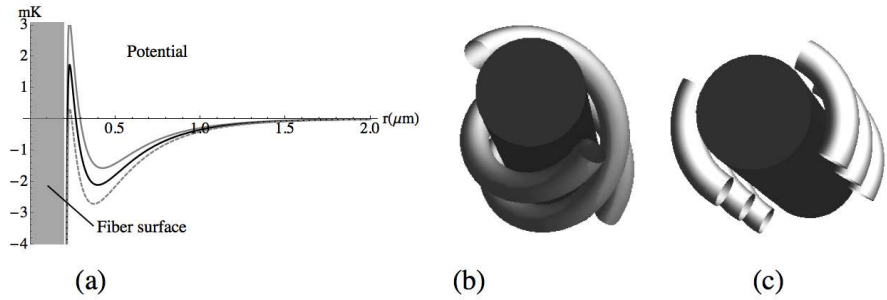}
   \caption{(a) Potential in the radial direction for a $200$nm fiber with $29$ mW in the blue-detuned beam and three different powers in
     the red-detuned beam: $25$mW (most shallow), $30$ mW (intermediate depth) and $35$ mW (deepest potential). Panels (b) and (c) show the   
     shape of the potential in three dimensions when the counter propagating red-detuned modes have parallel linear or orthogonal 
     circular polarization respectively.}
\label{fig:HE11Potential}
\end{figure}

A simple and stable ring trap in the transverse plane around the fiber can then be created  by combining a blue- and a red-detuned evanescent field, which due to their differing decay lengths results in an intensity minima at some radial distance from the fiber surface \cite{LeKien:04}.
However, the attractive red-detuned intensity must overcome the blue-detuned one at a position where the van der Waals potential is negligible. Though the van der Waals expression for an atom close to a curved dielectric surface is quite complex, in our scheme it is sufficient to approximate this attractive potential by the much simpler form for an infinite plane dielectric surface which is given by $V_{vdW}=- \frac{C_3}{(r-a)^3}$ with $C_3=5.6 \times 10^{-49}$Jm$^{3}$. This approximation is accurate for trapping minima close to the fiber surface \cite{LeKien:04}. The radial position of the  minimum can be adjusted by varying the powers in the red- and blue-detuned beams and in Fig.\ref{fig:HE11Potential} (a) we show typical examples for a fiber of radius $a=200$nm and two optical fields of $\lambda_{red}=1064$nm and $\lambda_{blue}=700$nm, which are red- and blue-detuned from the dominant D2 transition line in caesium at $\lambda = 852$nm . The fiber radius was chosen to ensure the single mode condition is satisfied for both wavelengths and the power in the blue beam is kept fixed. By increasing the power in the red beam, a clear shift of the minimum towards the fiber surface is observed.

It has recently been pointed out, by D. Reitz and A. Rauschenbeutel \cite{Reitz:12}, that a double helix potential can be constructed in the evanescent field of a single mode optical fiber by the appropriate combination of three circularly polarized light beams. Their proposal is based on the following principle; the intensity along the propagation ($z$) direction can be modified by using a counter-propagating beam configuration. If both beams are linearly polarized, then a simple standing wave pattern is formed. If the beams are circularly polarized, however, the standing wave formed by their superposition can take two different forms depending on whether the components have the same or opposite handedness. In the case where the counter-propagating components have the same handedness a standing wave with circular polarization and hence azimuthal symmetry in the intensity is formed. If the components have opposite handedness the standing wave is linearly polarized everywhere with the polarization direction rotating continuously through $2\pi$ along the propagation direction, resulting in the two intensity maxima which form a double helix pattern with a periodicity of $\lambda/2$ . The helical angle at the fiber surface is given by $\arctan(2\pi a/\lambda )$ \cite{Reitz:12}.

Figures \ref{fig:HE11Potential} (b) and (c) show the different shapes of such a potential in three dimensions for different polarisations
in the red-detuned beam. In (b) the counter propagating red-detuned beams have parallel linear polarisation and the resulting potential has the form of a double helix \cite{Reitz:12}. In (c) the beams are orthogonally circularly polarized and one observes the creation of a potential that is periodic in the azimuthal and longitudinal directions and has similarities to having two one-dimensional optical lattices aligned along opposite sides of the fiber. The trapping frequency in the radial and azimuthal direction can be determined from $\frac{\omega_r}{2\pi}=\sqrt{\frac{\partial^{2}_{r}U(r_{min})}{m}}$ where $m$ is the mass of caesium and one finds typical values of $500$KHz in the radial direction and $150$KHz in the azimuthal one.

\section {Trapping using the higher order $HE_{21}$ mode} 
\label{sec:Traps}

\begin{figure}[h]
   \centering
   \includegraphics[width=3.5in]{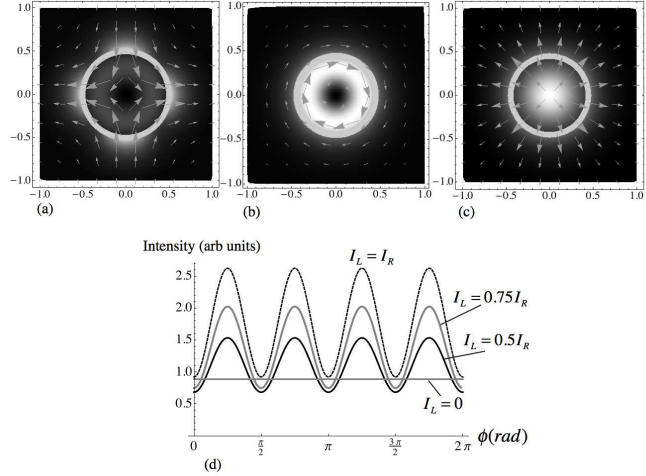} 
   \caption{ (a) Intensity and polarization vectors for the $HE_{21}$
     mode with quasi-linear polarization, for a $500$nm diameter fiber
     with $1064$nm wavelength light. (b) and (c) show the $TE_{01}$ and $TM_{01}$ 
     modes for the same parameters.  The ring marks the fiber
     vacuum boundary. (d) shows the intensity in the azimuthal direction for the $HE_{21}$ mode, at
     $100$nm from the fiber surface, for four different polarization
     states where $I_{L/R}$ denotes intensity in left/right circularly
     polarized modes.}
   \label{fig:HE21}
\end{figure}

If the parameter $V$ is increased beyond the value of 2.405, simply by increasing the fiber diameter, three additional modes will successively be allowed to propagate. These are the $HE_{21}$, $TE_{01}$ and $TM_{01}$ modes. In the weakly guiding approximation, all three of these modes correspond to the $LP_{11}$ mode which resembles the free space Laguerre Gaussian $LG_{01}$ mode. The strong confinement of the mode by the nanofiber causes the $LP_{11}$ mode to split into three modes with different propagation constants.The transverse intensity  and polarization profiles of these three modes are shown in Fig ~\ref{fig:HE21}. 

It is noteworthy that the $TM_{01}$ mode has significant on axis intensity, due to the large $z$-component of the field in the nanofiber, whereas the $TE_{01}$ and $HE_{21}$ modes have the same doughnut shape as the $LP_{11}$.  The transverse part of the polarization of the $TE_{01}$ and $TM_{01}$ modes is linear with a $2\pi$ rotation of polarization direction. 

Since trapping geometries based on superpositions of the $TE_{01}$ modes with both the $HE_{21}$ and the $HE_{11}$ modes are known to be very versatile \cite{Sague:08}, we present in the following an analysis of the trapping geometries that can be created by combining the $HE_{21}$ mode and the $HE_{11}$ fundamental mode. We will first consider a clean excitation and then include effects from unintended excitations of the $TE_{01}$ and $TM_{01}$ modes on these potentials. Even though these unintended excitations change the trapping potential, we show that this can actually lead to a new category of interesting geometries. 

The $HE_{21}$ mode can be either circularly or linearly polarized, where the linearly polarized one is a direct combination of right and left circularly polarized beams. In this case the transverse part of the polarization vector undergoes a $2\pi$ rotation in a circuit around the fiber axis, while the polarization vector rotates in the opposite sense to the circuit taken so that the electric field vector is orthogonal to the fiber surface at four points. The evanescent field must be continuous across the fiber boundary at these points and this results in four intensity maxima along the azimuthal direction. These polarization vectors are shown superimposed on the intensity profile at one instant of time in Fig ~\ref{fig:HE21}(a). The circularly polarized $HE_{21}$ mode has an azimuthally symmetric intensity profile (not shown).

 The variation in azimuthal intensity distribution for different polarization states of the $HE_{21}$ mode is shown in Fig.~\ref{fig:HE21}(d) and four intensity maxima are clearly visible, which become more pronounced the less circularly polarized the beam is. In these figures the fiber radius is $500$nm so that the $HE_{21}$ mode is
allowed to propagate for $1064$nm wavelength light in a silica fiber with a decay length of $1/q =0.52~\mu$m

\begin{figure}[tb]
   \centering
   \includegraphics[width=4in]{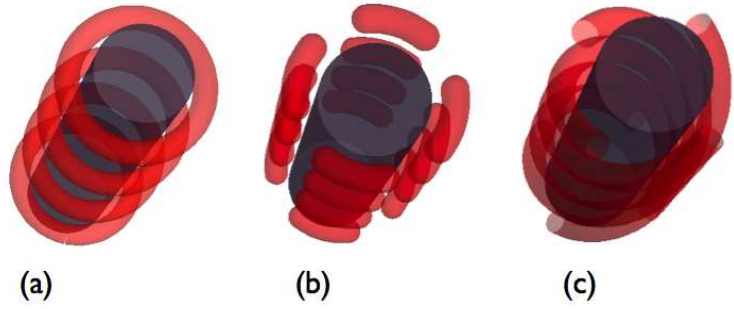} 
   \caption{Potential shapes corresponding to three types of standing wave (a) the shape of the potential in three dimensions
     when the modes have identical circular polarizations (b) the shape of the
     potential in three dimensions when the modes are linearly
     polarized, (c) the shape of the potential in three dimensions
     when the modes have orthogonal circular polarizations }
   \label{fig:HE21Potential}
\end{figure}

Counter-propagating $HE_{21}$ modes can broadly be achieved in three different ways, which lead to three different classes of standing wave pattern: 
(a) counter-propagating circularly polarized modes of the same handedness , 
{(b)} counter-propagating quasi-linearly polarized modes, 
and (c) counter-propagating circularly polarized modes of opposite handedness.
Two counter-propagating quasi-circularly polarized modes with the same handedness form a cylindrically symmetric standing wave (case {(a)}), which can be used to create symmetric,  equally spaced disconnected potential rings around the fiber, see Fig.~\ref{fig:HE21Potential}(a). 
Counter-propagating two quasi-linearly polarized $HE_{21}$ modes (case {(b)}), results in a standing wave with four intensity maxima per circuit of the fiber, which allows to create a regular lattice of traps with periodic boundary conditions around the fiber, see Fig.~\ref{fig:HE21Potential}(b). 
And two counter propagating $HE_{21}$ modes with opposite quasi-circular polarization (case {(c)}) will combine to form a linearly polarized mode like that shown in Fig ~\ref{fig:HE21}(c)  with the polarization vectors rotating with propagation direction. This results is an evanescent field consisting of four intertwined helices, leading to a potential of the same shape (see Fig.~\ref{fig:HE21Potential}(c)).  

\begin{figure}[tb]
   \centering
   \includegraphics[width=4in]{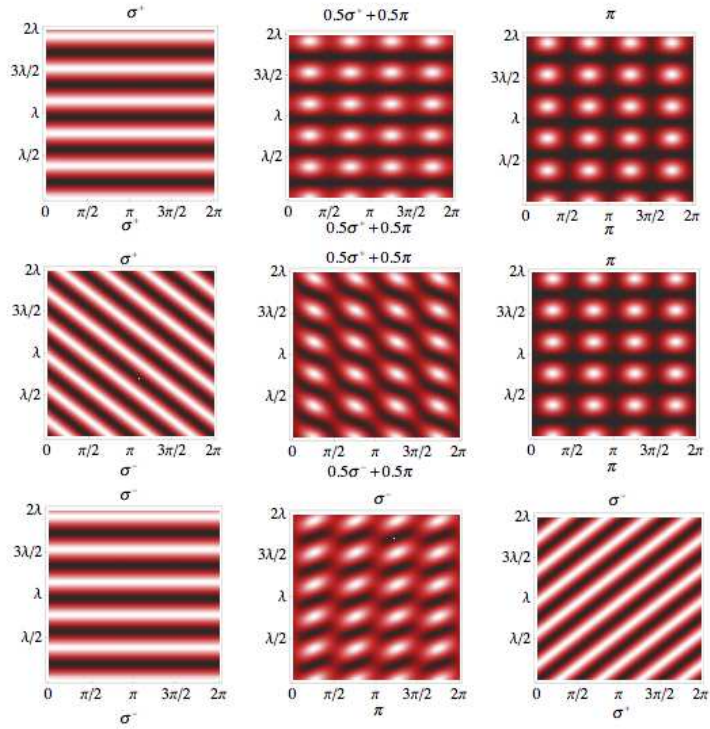} 
   \caption{Intensity in the $\{r,\phi\}$ plane as a circularly symmetric standing wave is transformed into the lattice (upper row), a four helix standing wave is transformed into the lattice (middle row) and a circularly symmetric standing wave is transformed into the four helix standing wave (lower row).}
   \label{fig:Transitions}
\end{figure}

It is experimentally conceivable, in principle, to switch between these different standing wave configurations by smoothly varying the polarization. In the upper row of Fig.~\ref{fig:Transitions} we show how the standing wave formed by two counter-propagating modes with the same quasi-circular polarization transforms into a two-dimensional lattice of traps by simply changing the polarization from circular to linear in each of the counter propagating modes simultaneously. The same method transforms the four helix standing wave into the lattice (see middle row of Fig.~\ref{fig:Transitions}) and in order to transform the four helix potential into the circularly symmetric standing wave one of the circularly polarized modes, say left circularly polarized, is held constant while the other is smoothly changed from right to left (lower row in Fig.~\ref{fig:Transitions}).

To use the above intensity patterns to trap caesium atoms, we consider introducing a $700$nm fundamental $HE_{11}$ mode, which, for the fiber parameters specified above, has an evanescent field decay length of $1/q_{blue} =0.12 \mu$m.  The three dimensional geometries of these potentials for counter-propagating modes with identical circular polarization, counter-propagating modes with parallel linear polarization and for
counter-propagating modes with orthogonal circular polarization are shown in in Figs.~\ref{fig:HE21Potential}(a),(b) and (c), respectively. The distance between potential minima in the azimuthal direction is
approximately a wavelength for the chosen parameters and depends on the radial
position of the potential minimum, which can be modified by varying the
ratio of power in the red-detuned field to that in the blue. Here the potential
minimum, for $26$ mW in the $1064$nm field and $110$ mW in the blue
field, is located at $145$nm from the fiber surface. 

Fitting the radial profile in the vicinity of the trapping minimum to a harmonic oscillator potential $\frac{1}{2} m_C
\omega_{r}^{2}r^2$, where $m_C$ is the mass of caesium, leads to an
estimate of $400$kHz for $\omega_r/2\pi$. The same method leads to an
estimate of $240$kHz for the azimuthal trapping frequency
$\omega_\phi/2\pi$. Note that the potential barriers between the traps can be
smoothly removed in the longitudinal direction by lowering the power
in one of the counter-propagating beams and independently in the azimuthal direction
by changing the polarization of the mode from linear to circular.

\section{Effects of Mode-Mixing}
\label{sec:powertransfer}

Controlling the distribution of power between the different modes allowed to propagate in a nanofiber is an experimentally open problem. Depending on the shape of the tapering region of the fiber and the quality of the beam injected initially, the modes present in the tapered waist can have different populations 
%, the diameter of the waist region plays a similar role determining the relative emission rates into the available modes
\cite{Masalov:13}. Simultaneous excitation of the $TE_{01}$, $TM_{01}$ and $HE_{21}$ modes has been demonstrated \cite{Petcu:11,Frawley:12}, however a clean excitation of the $HE_{21}$ mode has not yet been reported. We will therefore, in the following, look at the effects the presence of unintended excitation of the $TE_{01}$ and $TM_{01}$ modes has on the $HE_{21}$ based potentials.

\begin{figure}[H]
   \centering
   \includegraphics[width=4.5in]{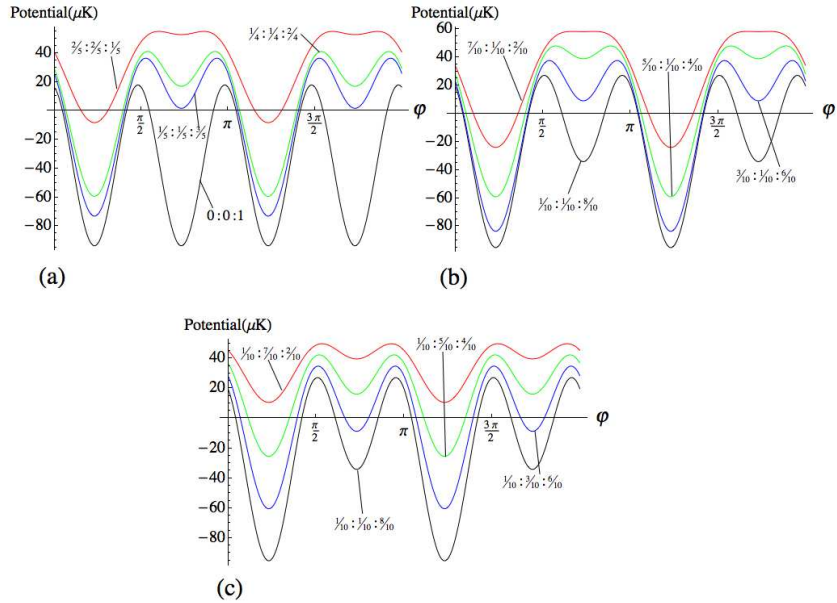} 
   \caption{ The modification of the 4-helix potential as power is transferred from $HE_{21}$ to (a)
    $TE_{01}$ and $TM_{01}$ in equal measure, (b) $TE_{01}$ with the power in $TM_{01}$ held constant and (c) $TM_{01}$ with $TE_{01}$ held constant. The ratios represent the fraction of power in each mode as $TE_{01}$:$TM_{01}$:$HE_{21}$}
   \label{fig:MixedModes}
\end{figure}

So far, our scheme has assumed that we can solely excite the $HE_{21}$ mode. However if power is passed from the $HE_{21}$ to the $TE_{01}$ and $TM_{01}$ modes, the 4-helix potential will be modified and in general the potential changes from having four traps per azimuth to only two (see Fig.~\ref{fig:MixedModes}). The effect is stronger in the case of power transfer to the $TE_{01}$ mode. 
 This can be explained by referring back to Fig.~\ref{fig:HE21} where it can be seen that the $TE_{01}$ and $TM_{01}$ modes can interfere destructively with the $HE_{21}$ mode at two azimuthal positions where the polarizations match. In the case of the radially polarized $TE_{01}$ mode this effect will be greater since the $HE_{21}$ evanescent field is strongest where the polarization is radial. In this figure there is $50mW$ propagating in the fundamental mode and $2mW$ shared between the three higher modes.
 
\section{Conclusion}
\label{sec:Conclusions}
By considering a tapered fiber with a diameter large enough to allow the next lowest modes above the fundamental mode to propagate, we have shown that a trapping potential in the shape of four intertwined helices can be created. This is achieved by combining red- and blue-detuned modes in a specific way and in particular by counter-propagating the red-detuned $HE_{21}$ modes. This 4-helix configuration can be transformed into a rectangular lattice of traps wrapped around the cylinder by adjusting the polarisations of the counter-propagating beams. The potential barriers between traps in the azimuthal and longitudinal directions can be controlled independently.  We have also considered the effects on this trapping geometry from unintended transfer of power from the $HE_{21}$ to the $TE_{01}$ and $TM_{01}$ modes. It is apparent that, were it possible to selectively couple to the available modes with strategically chosen relative intensities, a large number of interesting trapping geometries could be achieved. 

\section{Acknowledgements}
This work was supported by Science Foundation Ireland under project number 10/IN.1/I2979

\end{document}